\documentstyle[12pt,epsfig]{article}
\def \ins#1#2#3#4#5#6 {
  \begin{figure}[#1]
    \begin{center}
      \psfig{file=#2,width=#3,height=#4,angle=0}
      \caption{#5}
      \label{#6}
    \end{center}
  \end{figure}
  }

\begin{document}
\begin{center}
{\large \bf
Simulation of antiproton-nucleus interactions in the framework of
the UrQMD model}
\end{center}

\begin{center}
 A.S. Galoyan,  A.Polanski
\end{center}

\begin{minipage}{15cm}

This  paper  proposes  to apply the
 Ultra-Relativistic Quantum Molecular Dynamic (UrQMD) approach
to implement the PANDA  project (GSI).
 Modeling of $\bar{p}$A-interactions
has been performed
at antiproton energies from 1 GeV to 200 GeV by using the UrQMD model.
We have
studied  average multiplicities,  multiplicity distributions of
various types of secondary  particles, correlations between the
multiplicities,  rapidity, and transverse
momentum distributions of the particles.
The UrQMD model predictions on  inelastic $\bar{p}$A-
collisions have been found to
reproduce qualitatively the experimental data.  However, to
reach the quantitative agreement, especially, in fragmentation regions, it
is needed to modify the UrQMD model.

\end{minipage}

\section{Introduction}
In our paper we perform a comparative analysis of experimental
data on $\bar{p} A$-interactions with calculations using the well-known
Ultra-Relativistic Quantum    Molecular Dynamic model (UrQMD) \cite{URQMD}.

 The  property of
antinucleons - the ability to annihilate in nuclear matter, gives an
ample possibility to investigate  exotic nuclei and processes.
The presence of an annihilation channel in the $\bar{N} A$-interaction
leads to enhancement of such interesting effects in  intranuclear
cascades as multipion-nucleus interactions and the interaction of meson
resonance with nuclei \cite{Iljinov}. This also may result in
qualitatively new effects, for example,  the formation of a droplet
of "hot" nuclear matter \cite{Andreev5}, \cite{Andreev6}
 with its subsequent specific
decay. Specifically, in \cite{Miano}  it is shown that in
the interaction of antiprotons with tantalum nuclei
$K_s^0$-particles
are mainly produced from the cluster
containing p+3 nucleons with temperature $T\approx 135 $ MeV,
and $\Lambda $-particles --  from the cluster containing p+13
with temperature $T\approx 97$ MeV.

Analyzing  the annihilation of antinucleons in nuclei, it is
interesting, first of all, to determine the difference
of  the characteristics of secondary particles in the  $\bar{p} A$-
interactions from the predictions of microscopic models
for hadron-hadron  and nucleus-nucleus collisions, in particular,
 Ultra-Relativistic Quantum  Molecular Dynamic model (UrQMD),
and from the data on nucleon-nucleus interactions for which the UrQMD
model gives a qualitative agreement.
 The above mentioned
  specific features  may be associated with the
 effect of clustering  nucleons in the intranuclear annihilation
 process
 or, probably, some demerits of the model.

Our work has been inspired by the goals of the PANDA project at GSI.
Among them there are those which are especially
important for UrQMD contribution:
\begin{itemize}
\item Study of
interactions of hidden and open charm particles with nucleons and
nuclei;
\item Investigation of strange baryons in nuclei.
\end{itemize}
To reach these
goals, it is necessary
 to have $\bar{p} $-nucleus events code-generator
\begin{itemize}
\item[-] to estimate the background conditions for experiment;
\item[-] to develop the experimental setup, trigger conditions,
and so on.
\end{itemize}
Of course, the generator of $\bar{p} $-nucleus interactions
is needed to study the physics of  processes.

There are only few  event generators --
 FLUKA \cite{Fluka},  Intranuclear cascade model \cite{Golubeva}, DTUJET
 \cite{DTUJET}, UrQMD,
 to simulate $\bar{p} $-nucleus collisions.
   We assume that the UrQMD approach is the most suitable tool to solve
   these tasks.

\section{Main assumptions of UrQMD approach}

 A detailed model description can be found in  papers \cite{URQMD}.
 The UrQMD model is a microscopic transport theory based on the
 covariant propagation of all hadrons on classical trajectories
 in combination with stochastic binary scatterings, colour string
 formation and resonance decay.

 It represents a Monte-Carlo solution of a large set of coupled
 partial integro-differential equations for the time evolution of
 various phase space densities
 $f_i(x,p)$  of particle species
 $i=N,\Delta, \Lambda $, etc, that non-relativistically assumes
 the Boltzman form:
\begin{displaymath}
\frac{df_i(x,p)}{dt} \equiv \frac{\partial p}{\partial t}
\frac{\partial f_i(x,p)}{\partial p} + \frac{\partial x}{\partial t}
\frac{\partial f_i(x,p)}{\partial x} +
\frac{\partial f_i(x,p)}{\partial t}= St f_i(x,p),
 \end{displaymath}
 where
$ x $ and $ p $ are the position and momentum of particle, respectively,
$St f_i(x,p)$ denotes the collision term of these particle species,
which are connected to any other particle species  $f_k(x,p)$.

The UrQMD approach includes
\begin{itemize}
\item consideration of cross-section of various meson-meson,
meson-baryon, and baryon-baryon interactions.
It takes into account $\sim $ 50 baryons, $\sim $ 45 mesons.
\item It considers $Q \bar{Q} $ string creation a la FRITIOF model at
$P_{lab} >5 GeV/c $.
\item It also considers string fragmentation and formation time of
particles.
\item At lower energies, $P_{lab}<5$GeV/c there are  reaction channels:
\\ $ N + N \to \Delta N, \Delta \Delta, N^*N, etc.,$ $ M + N \to \Delta
^0, \Lambda,... $
\item  The  potential interactions are supposed between the
particles, especially,  Yukawa, Coulumb, Pauli potentials.
\end{itemize}

The physical picture of $hA$ collisions, used by UrQMD model
is approximately the following. The projectile hadron interacts with
the target nucleus nucleon.
A collision between two hadrons will occur if
$ d < \sqrt {\sigma _{tot} /\pi }$, where
$d$ and $\sigma _{tot} $ are the impact parameter of hadrons
and the total cross-section of two  hadrons, respectively.
In the UrQMD model the total cross-section  $\sigma _{tot} $
depends on the isospins of colliding particles, their flavor
and the c.m.s. energy. However, partial cross-sections
are then used to calculate the relative weights
for  different channels.
The total baryon-baryon  cross-section of the reaction
$ A+C \to D+E $ has the following general form:

$$ \sigma_{tot} ^{BB} \propto (2S_d + 1)(2S_E + 1)
\frac{\langle p_{D,E} \rangle }{\langle p_{A,C} \rangle }
\frac{1}{s}|M|^2, $$
with the spins of the particles, $S_i$, momenta of the pairs of
particles, $<p_{i,j}>$, in the two particle rest frame, and the matrix
element $|M|^2$.

At high energies,
two quark strings are produced
in the hadron interaction with target nucleon.
The first fast string
can  collide with other target nucleon or
emit a meson(s), before the collision with the second nucleon  of the
target.  The second  string, slow in the Lab. system, decays in the
nucleus and its products interact with the nuclear nucleons.  Here, the
creation of different meson and baryon resonanses are taken into
account. The resonances can interact with each other and nuclear
nucleons, before they leave the nucleus.
 As a result, a lot of particles are produced in the simulated
 $hA$-interactions with UrQMD  model.

 At high energies ($p_{lab}>2$ GeV/c)
baryon-baryon and meson-baryon cross-sections are
parameterized as:
$$\sigma_{tot,el}(p) = A + Bp^n + Cln^2(p) + Dln(p)$$
with  the parameters of the CERN-HERA fit \cite{15descr}, the
 laboratory momentum $p$ in GeV/c and the cross-section in mb.  To
 describe the total meson-meson reaction cross-sections,
the additive quark model (AQM) and the principle of detailed balance
 are used,
that assumes the reversibility of the particles interactions.  To
model meson-meson interactions above the resonance region
$\sqrt{s}>1.7$~GeV, one applies the rescaled total $\pi p$ cross
section:
$$\sigma_{tot} ^{MM}(\sqrt{s}>1.7GeV)=\sigma_{tot}^{\pi
 p}(\sqrt{s}) \frac{\sigma_{AQM}^{MM}}{\sigma_{AQM} ^{\pi p}}.$$

If we look at the  Cascade model \cite{cascade}, it only considers  the
reactions:
$$NN \to NN + m\pi , \pi N \to  \pi N,  \pi N \to N+ m\pi ,
 \pi + (NN) \to NN.$$
As you can see, there are no $\omega, \rho $-mesons,
$\Delta $-izobars, $\Lambda $-hyperons, etc. All of them are
presented in the  UrQMD approach.
As a result, the model UrQMD describes the hadron-hadron and
hadron-nucleus interactions quite well.
To test this, we  have considered  experimental data
on hadron-hadron and hadron-nucleus interactions.

\begin{figure}[cbth]
\psfig{file=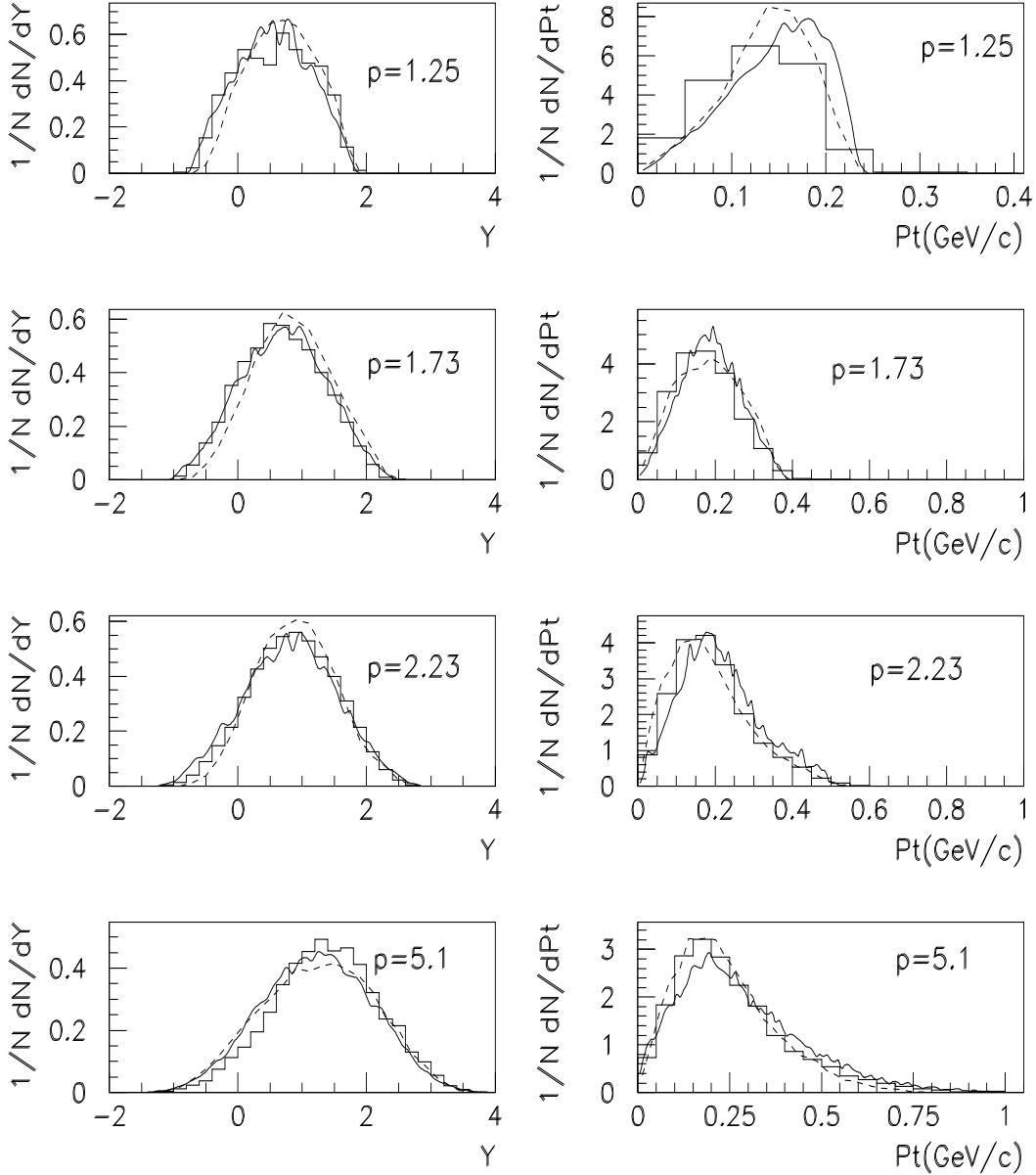,width=140mm,height=160mm,angle=0}
\caption{{\small $\pi ^-$meson
rapidity and transverse momentum distributions in np-interactions.
Histograms are experimental data
{\protect \cite{np}}.  Solid and  dashed lines are
UrQMD and modified FRITIOF {\protect \cite{Uzhinski}} model   calculations, respectively.}}
\label{pi1}
\end{figure}

 In  Fig.~\ref{pi1} we present
experimental data and UrQMD model calculations on $\pi ^-$-meson
rapidity and transverse momentum distributions in neutron-proton
interactions at the neutron momentum of $P_n=$1.25, 1.73, 2.23 and 5.1~GeV/c
 \cite{np}. The experimental data on  $np$-interactions were
obtained  with the 1-meter hydrogen chamber of
the Laboratory of High
 Energies (LHE), JINR,  by a group of Yu.A.~Troyan.
The histograms are experimental data, the solid lines are
calculations performed by means of the UrQMD model, the dashed lines are
modified  FRITIOF \cite{Uzhinski} model calculations.
As seen in  Fig.~\ref{pi1}, the qualitative and quantitative
description of $\pi ^-$-meson characteristics
was obtained by  the UrQMD model and the modified FRITIOF model.

\newpage

\begin{figure}[h]
\begin{center}
\vspace{-2cm}
\psfig{file=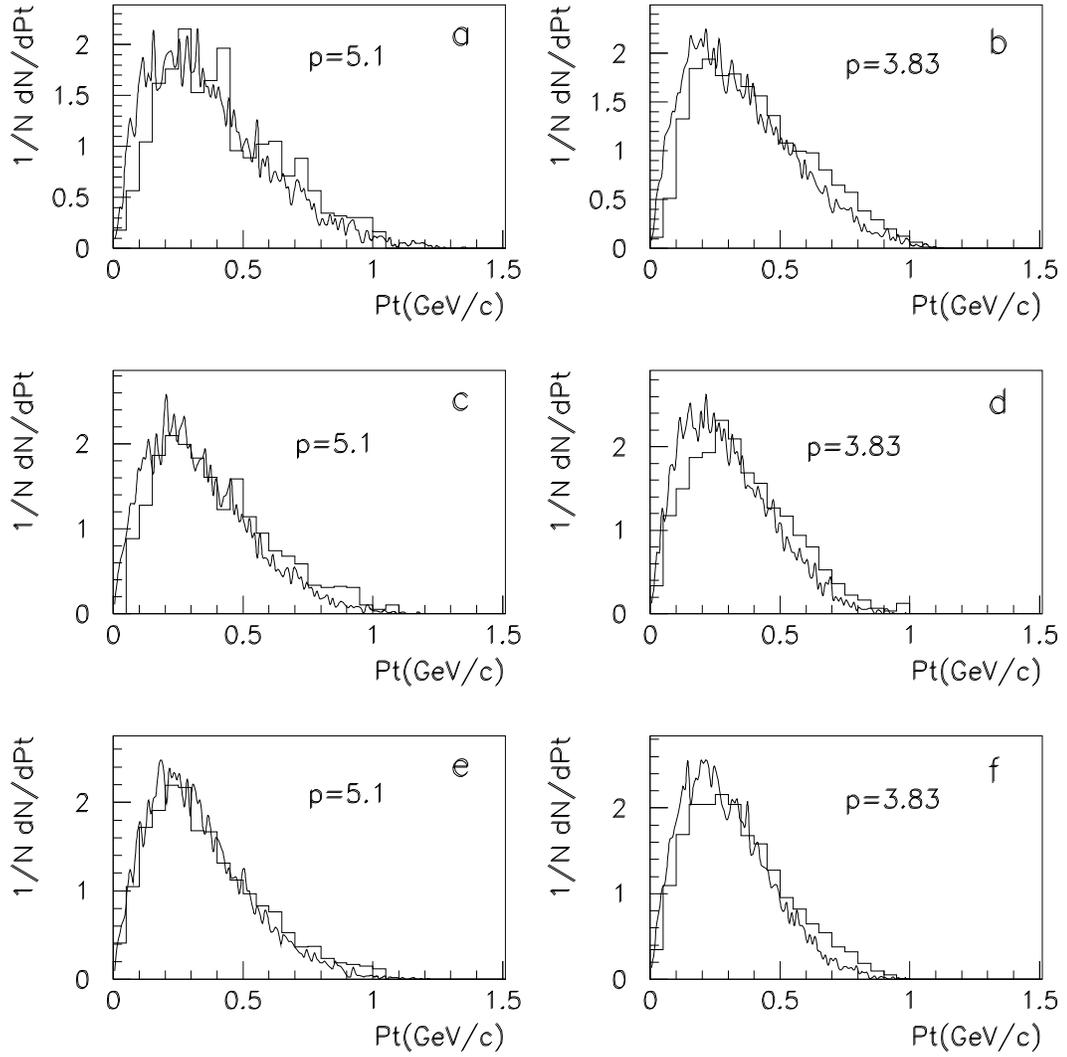,width=140mm,height=140mm,angle=0}
\vspace{-0.4cm}
\caption{{\small Proton transverse momentum distributions in
$np$-interactions.  Histograms are  experimental data {\protect \cite{np}}.
 Solid lines
are UrQMD model calculations.}}
\end{center}
\label{np2}
\end{figure}

Fig.~\ref{np2} gives  experimental and model distributions
of protons on transverse momentum in $np$-interactions in the reactions:
$np \to pp\pi ^- $ (Figs.~\ref{np2}a, \ref{np2}d),
$ np \to pp\pi ^-\pi ^0$ (Figs.~\ref{np2}b, \ref{np2}e),
 $ pp \to np\pi ^+\pi ^- $ (Figs.~\ref{np2}c, \ref{np2}f)
at the neutron momentum of $P_n=$3.83 GeV/c and 5.1 GeV/c.
Histograms are the experimental data, solid lines are the
UrQMD model calculations.
Fig.~\ref{np2} shows that the model predicts a larger yield of protons
at  small transverse momenta  and underestimates their number
at  large momenta.
However,
 the agreement of the calculations
of transverse momentum distributions of protons in
$np$-interactions
 with the
experimental data is as satisfactory, as  for $\pi^-$meson spectra.

\newpage
\vspace{-1cm}

\begin{figure}[cbth]
\begin{center}
\vspace{-3cm} \psfig{file=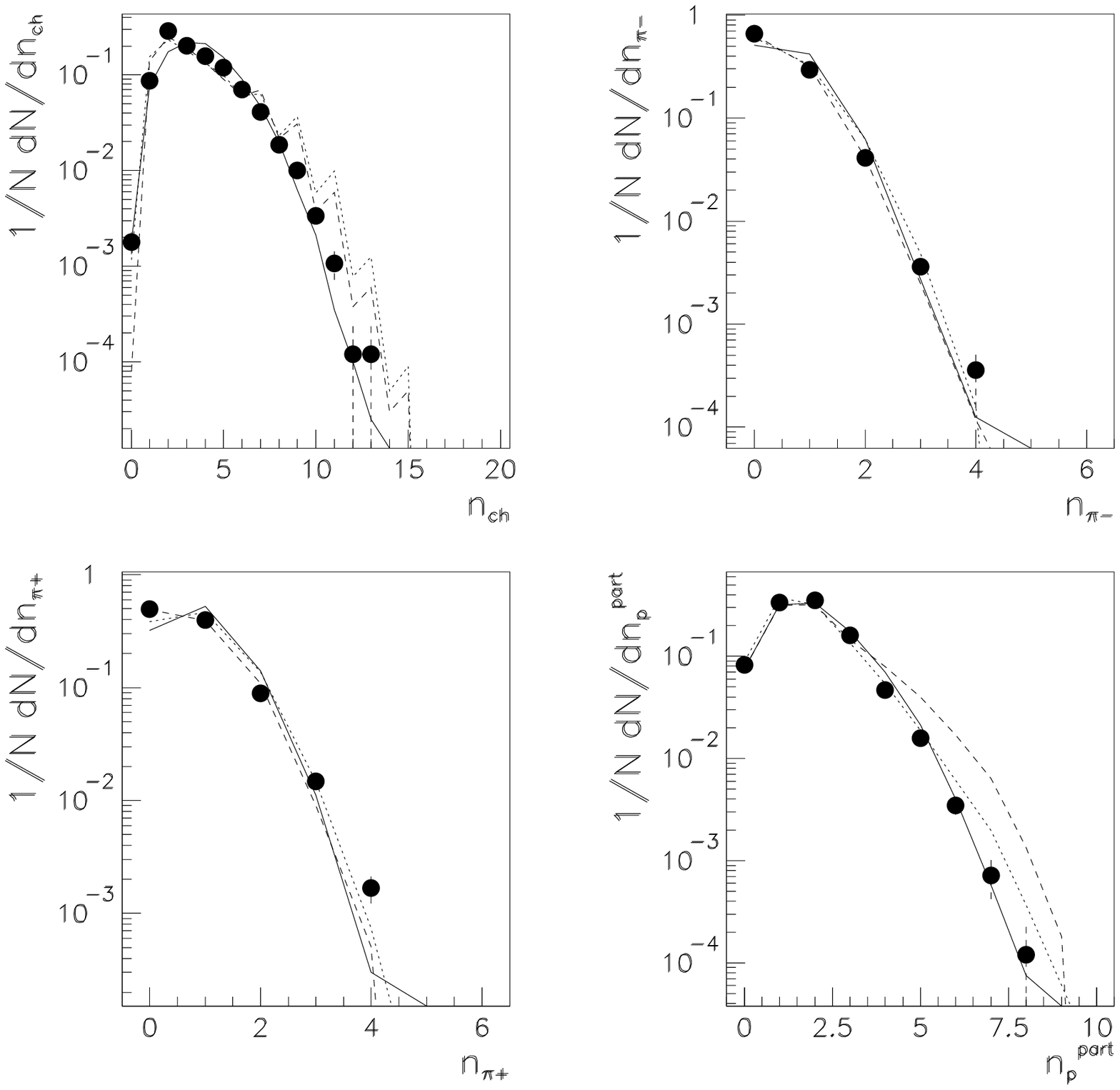,width=120mm,height=100mm,angle=0}
\vspace{-0.7cm}
\caption{{\small Secondary particles
 multiplicity distributions in pC-interactions.
Points are experimental data
\protect{\cite{bond}, \cite{pC}}.  Solid,  dashed and dotted lines are
UrQMD, modified FRITIOF \protect{\cite{pC}},
and cascade \protect{\cite{cascade}} model   calculations,
respectively.}}
\label{pc1}
\vspace{0.8cm} \psfig{file=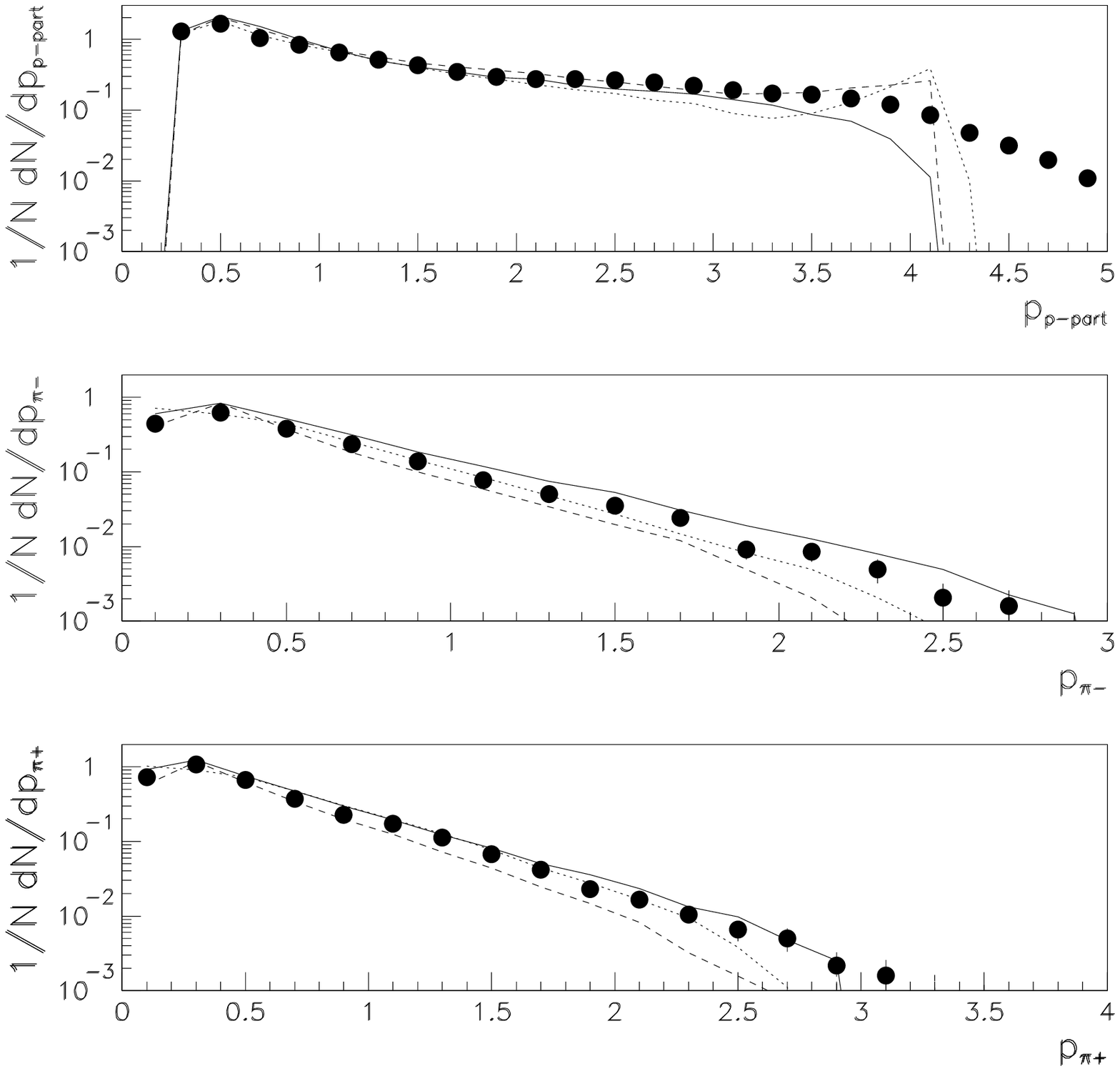,width=70mm,height=100mm,angle=0}
\vspace{0.8cm} \psfig{file=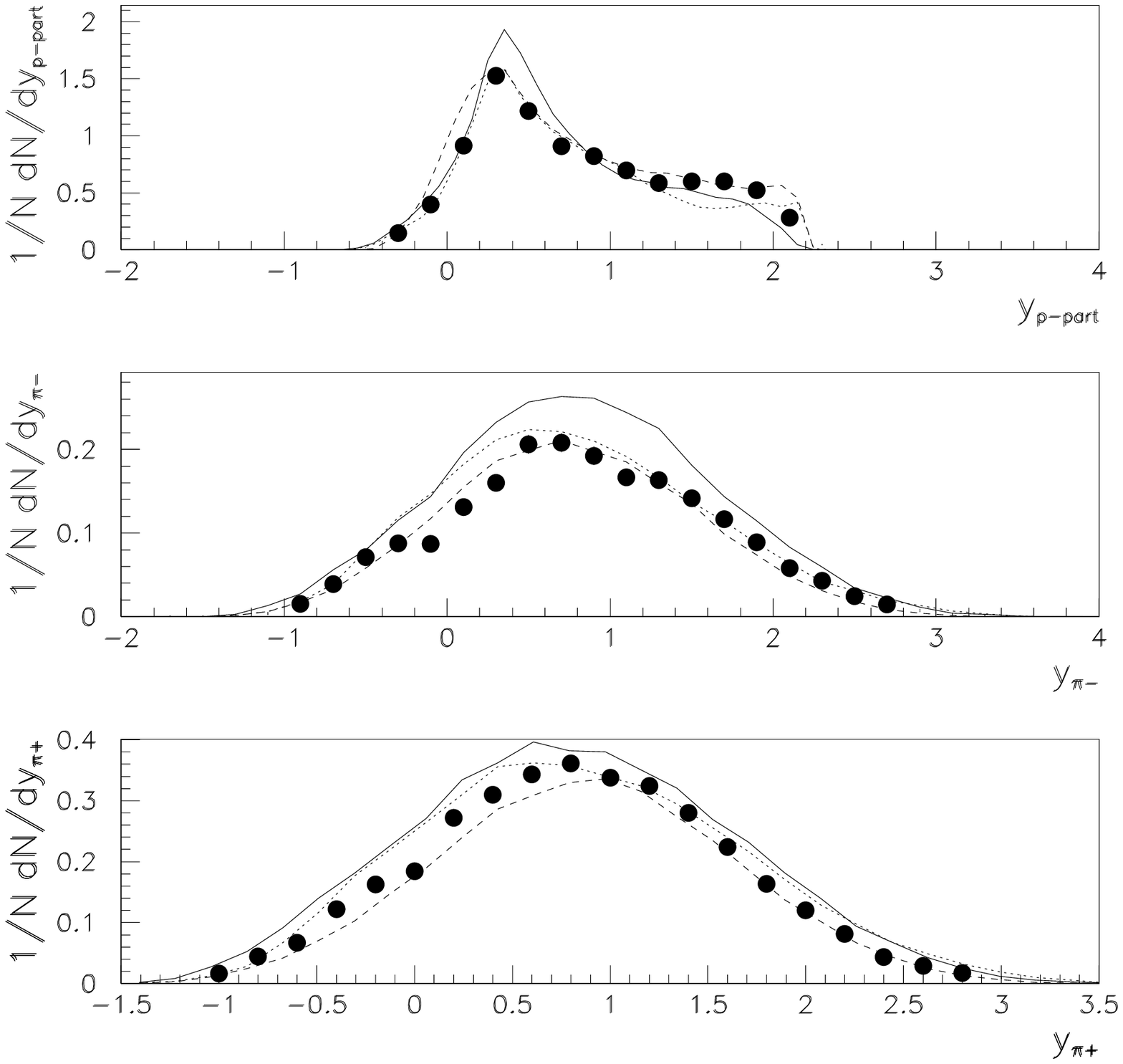,width=70mm,height=100mm,angle=0}
\caption{{\small Secondary particles
  distributions on momentum (left side), on rapidity (right side)  in pC-interactions.
Points are experimental data \protect{\cite{bond}, \cite{pC}}.
  Solid,  dashed and dotted lines are
UrQMD, modified FRITIOF \protect{\cite{pC}},
and cascade \protect{\cite{cascade}} model   calculations,
respectively.}}
\label{pc2}
\end{center}
\end{figure}

\vspace{-0.6cm}
 Figs. \ref{pc1}, \ref{pc2}
  present the data of $pC$-interactions at proton momentum
 of $P=$4.2~GeV/c, obtained by means of the 2-meter propan bubble
 chamber of LHE, JINR. A method of receiving and developing data is
 described in Ref. \cite{bond}, \cite{pC}.
Fig.~\ref{pc1} depicts the multiplicity distributions of
all charged particles,  $\pi ^-$ and $\pi ^+$ mesons,
and participant protons with the momentum more than 0.3 GeV/c
   in $pC$-interactions.
 The points are experimental data, the solid,
 dashed and dotted lines are the UrQMD, modified FRITIOF \cite{pC} and
 cascade \cite{cascade} model calculations, respectively.
 It is interesting, that the UrQMD model  describes the
 participant proton spectra better than the modified FRITIOF and the cascade model,
 and for the other groups of particles three models give the same good
 agreement with experimental multiplicity distributions.
 Fig.~\ref{pc2}  presents the kinematical characteristics
 of the secondary particles: participant protons with the momentum more
 than 0.3 GeV/c, $\pi ^-$ and $\pi ^+$ mesons.
  The distributions of secondary particles  on the  rapidity and  momentum are shown
 in  Fig.~\ref{pc2}.
   The UrQMD model
  calculation overestimates the production of $\pi $-mesons in the central
 rapidity region, and, especially, of the protons in the target fragmentation
region.  The same exceed of particle multiplicity  in the target fragmentation
region takes place in $\bar{p}A$-nucleus interactions, as it will be shown below.
As seen in  Fig.~\ref{pc2}, the UrQMD model calculations  reproduce
 quantitatively the
experimental momentum distributions of $\pi$-mesons.
On the whole, the UrQMD model calculations are in agreement with
experimental data of the $pC$-interactions, as it was  expected.
\vspace{-3cm}

\section{Results for $\bar{p}A$-interactions  and analysis}
Antibaryon-baryon interactions
are especially considered in the UrQMD approach.
It is well-known that at the energies $ \leq 100$~GeV/c
an important contribution to the total interaction cross-section
comes from the  annihilation  process , where only mesons remain in
the final state.
In the model, the $\bar{p}p$ annihilation cross-section is
parameterized by the relation taken from Koch and Dover \cite{26 UrQMD}
$$
\sigma ^{\bar{p}p}_{ann} = \sigma ^N_0 \frac{s_0}{s} \left[ \frac{A^2
s_0}{(s-s_0)^2+A^2 s_0} + B \right] $$
with $\sigma ^N_0$=120 mb,
$s_0=4m^2_N$, $A=50$MeV and $B=0.6$.
As known, the $\bar{n}p$ cross-section does
not differ significantly from $\bar{p}p$ cross-section \cite{27 UrQMD},
hence, they are set equal in the UrQMD model.  It is assumed, that two
massive strings and one meson
creation take place at the annihilation.  Since the
strings have practically the same energies and masses and are flying in the
same direction, we can expect, that there is a double multiplicity
of produced particles, compared with the multiplicity in
$NN$-interactions.  Since the interference between the particles is
not considered in the UrQMD model, we can expect that there is a
more powerful cascade compared with the cascade in hadron-nucleus
interactions.

\begin{figure}[cbth]
\begin{center}
\vspace{-0.3cm}\psfig{file=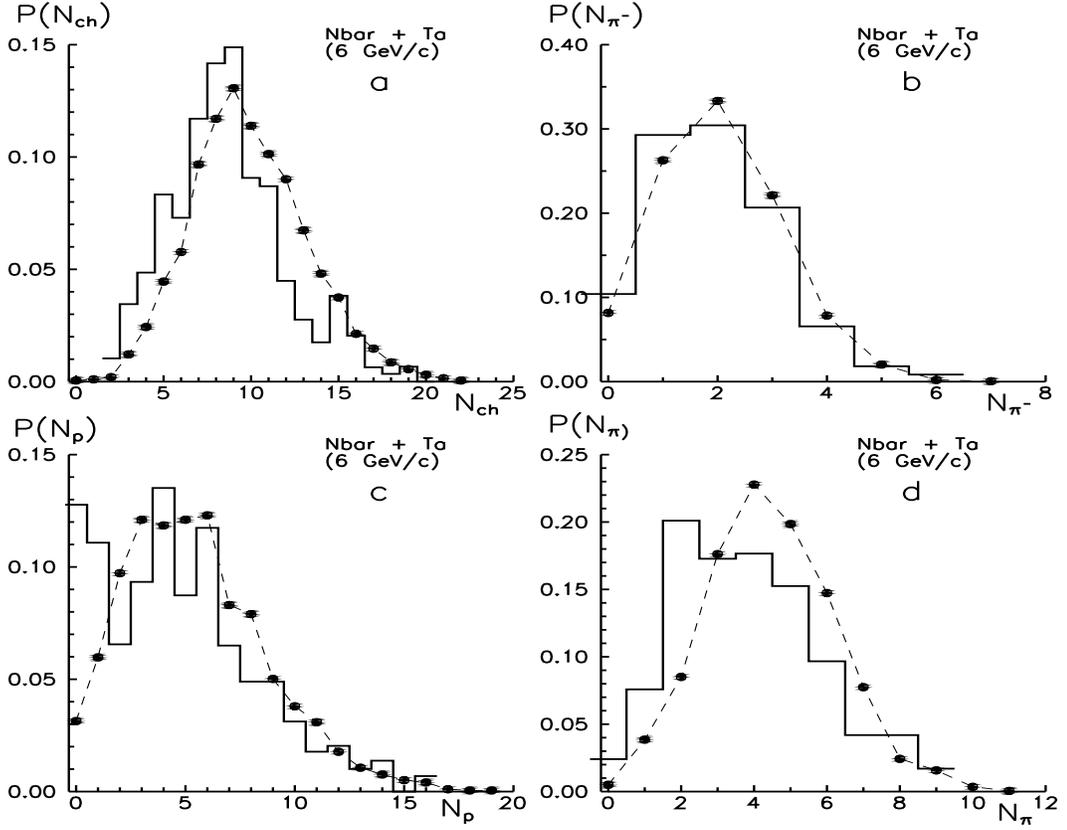,width=140mm,height=110mm,angle=0}
\caption{{\small Multiplicity distributions of secondary particles:
a) all charged particles,$N_ch$,
b) negative charged particles, $N_\pi -$,
c) identified protons, $N_p$,
d) charged mesons, $N_\pi$.
Histograms are experimental data
{\protect \cite{Andreev}}, points and dashed lines are UrQMD
model calculations.}}
\label{fig6}
\end{center}
\end{figure}

To check the expectation, we turn to the experimental data on $\bar{p},
\bar{n}$-interactions with nuclei. In the well-known "Guide to
experimental particle physics literature" \cite{1guide}, \cite{2guide},
there are only few references on antibaryon-nucleus interactions
 in the energy range from 1 to 20 GeV/c \cite{Andreev}, \cite{Jahnke}.
However, there are a lot of references on $\bar{p}A$-interactions at
energies 40~GeV/c, 100~GeV/c, 200~GeV/c.

In  Fig.~\ref{fig6} we present the  charged particles multiplicity
distributions in $\bar{n}Ta$-interactions at 6~GeV/c.
The experimental data
\cite{Andreev} were obtained by using the two-meter liquid-hydrogen
chamber "LYUDMILA" of the LHE of the JINR.  A
tantalum target was placed in the active volume of the chamber, which
was exposed to the beam of antideutrons at momentum 12.2~GeV/c
in the accelerator of the Institute for High-Energy Physics at
Serpukhov.    The statistics of experimental data
consists of 250 events. Our statistics includes 2000 events.
 Fig.~\ref{fig6}a shows the multiplicity distribution of all charged particles, $N_{ch}$.
 In the experiment \cite{Andreev}, the protons  with momentum less than
 1~GeV/c have been identified.
   Fig.~\ref{fig6}c gives the multiplicity distribution of slow (identified)
protons, $N_p$.
 Figs.~\ref{fig6}b and \ref{fig6}d illustrate the multiplicity distribution of
secondary negative charged particles, $N_{\pi -}$,
 and of all charged mesons,$N_\pi$,  respectively.
 These so-called  mesons (Fig.~\ref{fig6}d) include
 all charged mesons and unidentified protons with momentum more than
1~GeV/c.
As seen, there is a qualitative agreement between the
experimental data and the theory.

\begin{table}[h]
\begin{tabular}{|l|l|l|l|l|l|l|}
 \hline
 & $N_{event}$ & $<N_{ch}>$ & $<N_p>$ & $<N_{\pi}>$ & $<N_{\pi -}>$ &
 $<N_{\pi +} >$
 \\ \hline
 $\bar{n}Ta$ & 209 & 8.91 $\pm $ 0.24 & 4.93
$\pm $ 0.26 &  3.92 $\pm $0.14 & 1.97 $\pm $ 0.09 & 1.96 $\pm $ 0.09 \\
 \hline
UrQMD & 2000 & 9.896 &  5.452 & 4.444 & 2.024 & 2.420
 \\ \hline
ICM & 1000 & 8.12 & 2.91 & 5.21 & 2.57 & --
\\ \hline
\end{tabular}
\caption{{\small Average multiplicities of secondary charged
particles  in $nTa$-interactions at $P_{lab}=6$GeV/c}}
\end{table}

Table 1 lists the considered experimental data \cite{Andreev} (the first line)
 and calculations
of the  average multiplicities of secondary charged particles
for the interactions of antineutrons with tantalum nuclei at 6 GeV/c.
We present the simulations in the framework of the
UrqMD model (second line) and the intranuclear cascade model (ICM)
\cite{Golubeva} (third line).
The  average multiplicity
of all charged particles, $N_{ch}$,  calculated with the UrQMD model,
overestimates the experimental one by 10$\%$.
The theoretical average  multiplicity  of identified
protons, $N_p$, also exceeds the experimental one by 10$\%$,
the calculated multiplicity of the so-called $\pi $ mesons,
$N_{\pi }$,
overestimates the experimental data by 13$\%$, the simulated average
multiplicity of negative charged particles, $N_{\pi ^-}$,
 differs from the experimental
multiplicity only by 3$\%$. The big difference is observed between
the experimental data   and
 UrQMD calculations  on positive mesons and
unidentified protons, $N_{\pi ^+}$, it reaches 22$\%$.
 Table~1 illustrates the big discrepancy in the intranuclear cascade
 model calculations and experimental data, especially, for average
 multiplicity of identified protons.  This difference equals
 $\sim 40 \% $.

\begin{figure}[cbth]
\begin{center}
\vspace{-0.4cm}
\psfig{file=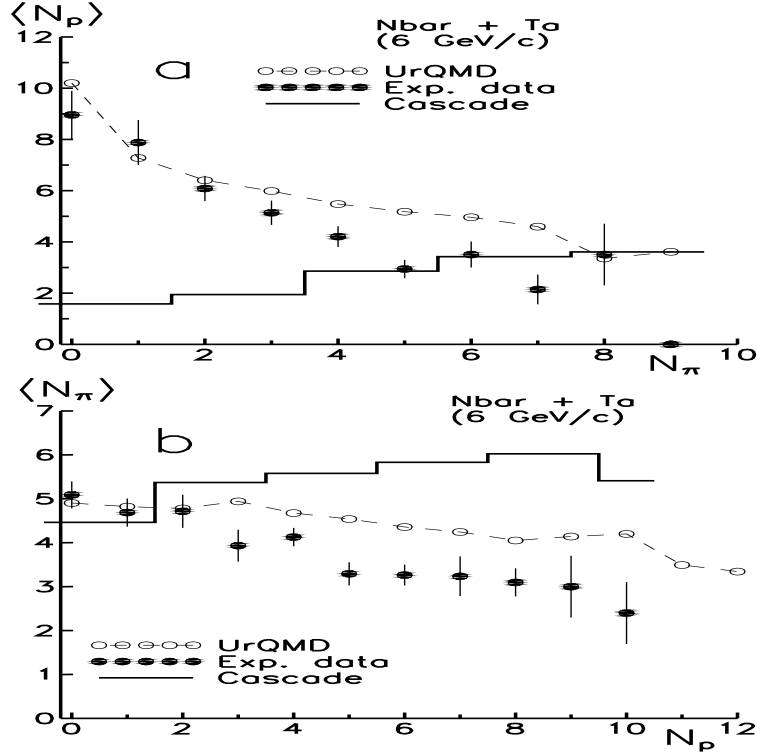,width=100mm,height=100mm,angle=0}
\vspace{-0.4cm}
\caption{{\small Correlations between protons and pions:
a) dependence of the average number of
 protons, $<N_p>$, on the number of pions,$N_\pi$,
b) dependence of the average number of pions, $<N_\pi>$, on
the number of identified protons, $N_p$.
Filled circles are experimental data
{\protect \cite{Andreev}}, dashed lines
and histograms are UrQMD and ICM \protect{\cite{Golubeva}} calculations,
respectively.}}
\label{fig8}
\end{center}
\end{figure}

 The correlations between multiplicities of mesons and identified
 protons give additional information on features of
 $\bar{n}Ta$-interactions.  Fig.~\ref{fig8}
 displays various experimental correlations between protons and pions
 with the corresponding UrQMD and ICM calculations.
  From Fig.\ref{fig8}a one can see that the experimental average multiplicity
 of the identified protons (solid points)
  falls with increase of the pion production
 by $\sim $  2.5 - 3  times.
  The average multiplicity of $\pi$-mesons   decreases
 with enhance of the   identified protons yield (see
 Fig.\ref{fig8}b)  by $\sim $ two times.   On the whole, the
 UrQMD calculations (dashed lines and open points) reproduce quite well
   the shape of the experimental correlations, but there is a
 quantitative discrepancy.  At the same time, predictions of ICM (the
 histograms) differ very strongly from the experimental data (see
 Figs.~\ref{fig8}a, \ref{fig8}b).  Due to low statistics of this
experimental data  we cannot make
 a  conclusion on the accuracy of calculations with the UrQMD
 model.

\vspace{-0.3cm}

\begin{figure}[cbth]
\begin{center}
\vspace{-0.6cm}
\psfig{file=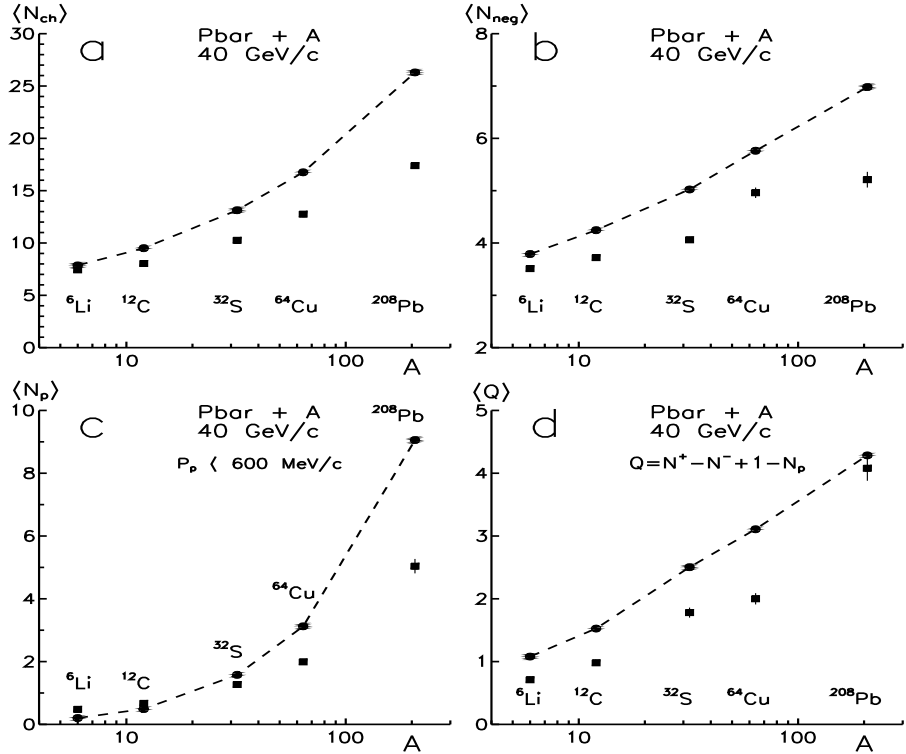,width=120mm,height=100mm,angle=0}
\vspace{-0.4cm}
\caption{{\small Average multiplicities of secondary  particles:
a) all charged particles, $<N_{ch}>$,
b) negative charged
particles, $<N_{neg}>$,
c) identified protons, $<N_p>$,
d) fast protons, $<Q>$,  vs atomic weight.
Squares are  experimental data \protect{\cite{Boos}}.
Dashed lines and circles are UrQMD model calculations.}}
\label{fig9}
\end{center}
\end{figure}

\vspace{-0.6cm}
Let us look at the  experimental data \cite{Boos}, \cite{Gabunia}
on $\bar{p} +Li, C, S, Cu, Pb$ interactions at
$\bar{p}$   momentum of 40~GeV/c.
These interactions were studied with
the RISK-streamer chamber spectrometer. In  Fig.~\ref{fig9}, the average
multiplicities of all charged particles, $~<~N_{ch}>$,  the negative
charged particles, $<N_{neg}>$,  the slow, identified protons with
momentum less than 500~MeV/c, $<N_p>$, and fast protons, $<Q>$, are
given as  functions of the atomic weight $A$. Since one cannot
 determine the number of fast protons exactly in the  experiment
 \cite{Boos}, therefore, the multiplicity of the so-called fast protons is
defined as:  $ Q= N^+ - N^- + 1 - N_p $, where $N^{+(-)}$ is the number of
positive (negative) particles.
 For experimental data (square  points),
Fig~\ref{fig9}b  shows that the
 increase with $A$ of the average number of slow  (identified)
protons  is much stronger than the linear behaviour and can be
parameterized as $<N_p> =c \cdot A^\alpha $, where $\alpha \approx
0.62$. The high value of $\alpha $ indicates that the slow protons
are governed by cascading and, probably, by an admixture of some evaporation
protons (at the $10\%$ level  \cite{Boos} ).
 From Fig. \ref{fig9}d, it is obvious that the average number
of fast protons, $<Q>$, depends much weaker on A than $<N_p>$.
The overestimation of the  UrQMD model simulations
(dashed lines in Fig.~\ref{fig9}) on the
average multiplicities of all charged particles, grows while increasing
the atomic weight $A$.
This is also true for the  negative particles average multiplicity,
$<N^->$, and the mean number of slow  protons, $<N_p>$.  This exceeding can
be explained by the larger cascading of intermediate particles in
nuclear matter on the model UrQMD  than it takes place in the reality.

\begin{small}
\begin{table}[h]
\vspace{-0.7cm}
\begin{tabular}{|l|ll|l|l|l|l|l|}
 \hline
 {\small Reaction} &  & $N_{event}$ & $ <N_p>$ & $ <N_{pb}>$  & $ <Q>$ & $ <N_{neg}>$
&  $<N_{ch}>$ \\
\hline
$\bar{p}Li$ & EXP & 1156 & $0.48\pm0.03$ & $ 0.14\pm0.02$ &
$0.71\pm0.04$ & $ 3.51\pm0.05$  & 7.43$\pm$0.12 \\
& UrQMD & 3649 &  0.202  & 0.08   &  1.08 & 3.79 & 7.86
  \\ \hline
$\bar{p}C$ & EXP & 1113 & $0.67\pm0.04$ & $0.24\pm0.02$ &
  $0.98\pm0.05$ & $3.72\pm0.10$ & 8.04$\pm$0.14
  \\
  & UrQMD & 4036   &  0.489  & 0.2 & 1.53  &  4.24  &  9.5
 \\ \hline
 $\bar{p}S$ & EXP & 1135 & $1.27\pm0.07$ &
$0.47\pm0.03$ & $1.78\pm0.08$ & $4.06\pm0.10$ & 10.26$\pm$0.18\\
& UrQMD & 4446  & 1.58 & 0.66 &  2.5  &  5.02 &  13.13
 \\ \hline
$\bar{p}Cu$ & EXP & 1419 & $1.99\pm0.09$ &
$0.64\pm0.04$ & $2.00\pm0.09$ & $4.96\pm0.10$ & 12.76$\pm$0.18 \\
& UrQMD & 4647 & 3.12 & 1.29 &  3.11 & 5.76 & 16.76
\\ \hline
$\bar{p}Pb$ &   EXP &   1108 & $5.04\pm0.23$ & $1.89\pm0.11$ &
$4.08\pm0.20$ & $5.21\pm0.15$ & 17.39$\pm$0.32
\\
& UrQMD & 3308  & 9.06 & 3.67  & 4.28 &  6.98 &  26.3
\\ \hline
 \end{tabular}
\vspace{-0.4cm}\caption{{\small Average multiplicities of secondary charged particles
 at $P_{lab}=40$GeV/c }}
\end{table}
\end{small}

\vspace{-0.2cm}
Table~2 shows
that the disagreement reaches 60$\%$ for $<N_{ch}>$, 40$\%$ for
$<N^->$, and 80$\%$ for  $<N_p>$ for $\bar{p} $ interaction with lead
 nucleus.  The UrQMD model calculations considerably exceeds the
average multiplicity of the identified protons, emitted in backward
 hemisphere,  $<N_{pb}>$, in Lab. system \cite{Gabunia} for medium and
 heavy nuclei.

Let us consider experimental data \cite{whitmore} on $\bar{p}A$-interactions
at momentum of anytiprotons 100~GeV/c.
This experiment was performed with the Fermilab 30-inch bubble chamber and
Downstream Particle Identifier to study inclusive charged pion production in the
high energy interactions of $\pi^{\pm }, K^{\pm }, p $ and $\bar{p}$ with
thin foils of magnesium, silver and gold.
 The laboratory rapidity and transverse momentum distributions are
 presented separately
for $\pi^+$ and $\pi^-$ production  and
compared with the  VENUS string model Monte Carlo in Ref. \cite{whitmore}.
Laboratory rapidity is defined as
$$ y = \frac{1}{2}  ln \left(\frac{E+P_z}{E-P_z}\right ), $$
where $E$ and $P_z$ are the laboratory energy and longitudinal momentum,
respectively.

\begin{figure}[cbth]
\begin{center}
\vspace{-0.5cm}\psfig{file=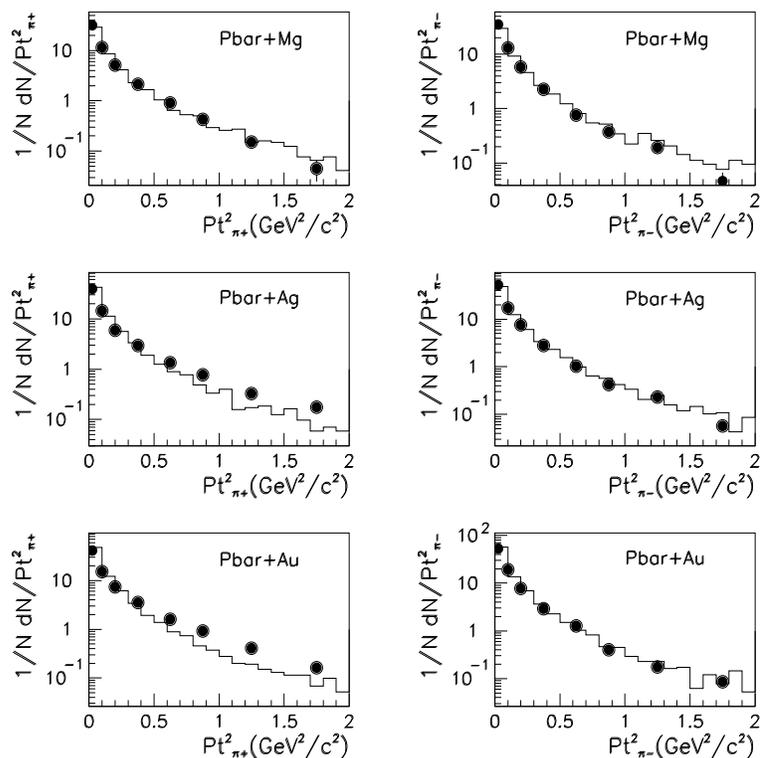,width=100mm,height=100mm,angle=0}
\vspace{-0.5cm}
\caption{{\small $P_t^2$ distributions of $\pi ^+$ and $\pi ^-$ produced
in  $\bar{p}A$-interactions at antiproton momentum 100~GeV/c.
Solid points are  experimental data \protect{\cite{whitmore}}.
Histograms are UrQMD model calculations.}}
\label{whit1}
\end{center}
\end{figure}

\begin{figure}[h]
\begin{center}
\vspace{-0.3cm}
\psfig{file=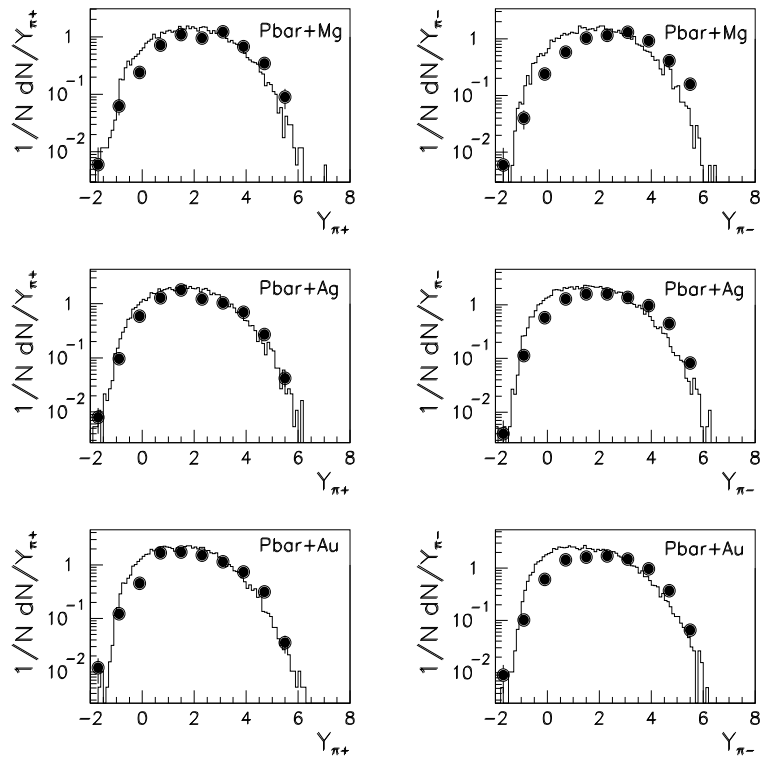,width=100mm,height=100mm,angle=0}
\vspace{-0.4cm}
\caption{{\small Rapidity distributions of $\pi ^+$ and $\pi ^-$ produced
in  $\bar{p}A$-interactions at antiproton momentum 100~GeV/c.
Solid points are  experimental data \protect{\cite{whitmore}}.
Histograms are UrQMD model calculations.}}
\label{whit2}
\end{center}
\end{figure}

We have performed the   modeling of the reactions
$$\bar{p} + Mg,~Ag,~Au \to  \pi ^{\pm } + X ~~~~~~~~~~(1)$$
at momentum 100 GeV/c
in the framework of the UrQMD model.  Figs.~\ref{whit1}, \ref{whit2} show
the results of the modeling and the corresponding experimental data \cite{whitmore}.

\vspace{-0.5cm}
Fig.~\ref{whit1} illustrates the transverse momentum distributions
of $\pi ^+$ and $\pi ^-$ - mesons.
All of the spectra are similar in shape, but the $Au$ data have slightly higher
cross sections due to the higher charged particle
multiplicities (see Table 3), and the $Mg$ data have the smallest cross sections.
All of the data show distributions which are inconsistent with a single
exponential slope but could be fitted with the sum of two exponential distributions.
 The UrQMD model calculations
reproduce the experimental transverse momentum distributions of pions,
even in the regions of large $P_t$,
which  are not described correctly with  most of Monte Carlo models.
In Fig.~\ref{whit2} we plot the pion rapidity distributions for
all reactions (1).
 All of the plots are similar and display the following features.
 We  observe that in the backward direction ($y<2$) they rise rapidly with increasing A.
As the rapidity increases, the $A$ dependence becomes weaker and in the region
 $y \approx 4.5$ the distributions become almost independent of $A$.
 We also do not observe  any region where a rapidity plateau exists.

\vspace{-0.3cm}

\vspace{-0.4cm}
The  UrQMD model  simulations  shift (a little bit)  the rapidity distributions
of   $\pi ^+$ and $\pi ^-$- mesons to the target
fragmentation region. On the whole,
the agreement between UrQMD  and data \cite{whitmore} is reasonably good.
Let us note, that the
UrQMD model describes   the $\pi ^+$ and $\pi ^-$ kinematical spectra better
than  the VENUS model.
In particular, the Venus model calculations
strongly underestimate the yield of pions in the central rapidity region,
and decrease essentially the meson production at large $P_t$ (see \cite{whitmore}).

\vspace{-0.2cm}
\begin{table}[h]
\begin{center}
\vspace{-0.3cm}
\begin{tabular}{|l|ll|l|l|}
 \hline
 {\small reaction} &  & $N_{event}$ &  $<N_{\pi +}>$ & $<N_{\pi -} >$
 \\ \hline
 $\bar{p}Mg$ & EXP & 218 & 4.30 $\pm $ 0.19 &  4.68 $\pm $ 0.16
\\
            & UrQMD & 1701  &  5.12  & 5.49
 \\ \hline
$\bar{p}Ag$ & EXP & 582 & 5.63 $\pm$ 0.15 & 6.41 $\pm$ 0.11
\\
         & UrQMD & 1841 &   7.05 &   7.96
\\ \hline
$\bar{p}Au$ & EXP & 465 & 6.24 $\pm$ 0.16 & 6.71 $\pm$ 0.13
\\
        & UrQMD & 1906 & 7.8 &    9.09
\\ \hline
\end{tabular}
\caption{{\small Average multiplicities of secondary charged
particles  at $P_{lab}=100$GeV/c}}
\end{center}
\end{table}

\vspace{-0.3cm}
However, the average multiplicity of $\pi ^+$ and $\pi ^-$ - mesons
produced in $\bar{p}A$-interactions at 100~GeV/c
calculated with UrQMD considerably
overestimates the  experimental ones, as it was observed
for the reactions $\bar{p}A$ at 40~GeV/c.

 Table~3  presents the experimental and  the UrQMD model calculated
average multiplicity of $\pi ^+$, $<N_{\pi ^+}>$, and $\pi ^-$,
$<N_{\pi ^-}>$, produced
in the reactions $\bar{p} + Mg,~Ag,~Au $ at momentum of antiprotons 100 GeV/c.
For $Mg$ and $Ag$ targets, the difference between the UrQMD calculations
and experimental results is equal to $\sim 20\% - 25\% $,
and $\sim 30\% - 35\% $ for $Au$ target.

A similar situation with  the multiparticle production
 in the $\bar{p}A$-interactions takes place at  momentum 200~GeV/c.
  Table~4 lists  experimental data \cite{De Marzo}
  and calculations by the UrQMD approach on
  the number of events,  the average multiplicities of all
charged particles, $<N_{ch}>$,
 the  negative  particles, $<N_{neg}>$,
 the forward emitted  particles in the CMS system, $<N_F>$,
and   the backward  ones, $<N_B>$ for $ \bar{p}p,~~\bar{p}Ar,
~~\bar{p}Xe $ interactions at the 200~GeV/c  antiproton beam.

\begin{table}[h]
\begin{tabular}{|l|ll|l|l|l|l|}
 \hline
 {\small Reaction} &  &
  $N_{event}$ & $<N_{ch}>$ & $ <N_{neg}>$ & $<N_F>$ & $<N_B>$  \\ \hline
$\bar{p}p$ & Exp. & 1856 &  7.69 $\pm$0.09 & 3.90 $\pm$ 0.05 &
3.97$\pm$0.06 & 3.56$\pm$0.06 \\
& UrQMD & 5000 & 6.009 & 3.003 &  3.277 &   2.732
\\ \hline
$\bar{p}Ar$ & Exp. & 577 & 16.01$\pm$0.40  & 6.80$\pm$ 0.16 &
5.45$\pm$0.12 & 8.97$\pm$0.29 \\
& UrQMD & 1896 & 19.243 &  8.083 &   5.205 &   14.038
\\ \hline
$\bar{p}Xe$ & Exp. & 905 & 22.41$\pm$0.46 & 8.54$\pm$0.15 &
6.19$\pm$0.10  &  12.75$\pm$0.33 \\
& UrQMD & 4694 & 29.061 &   10.386 &   5.642  &   23.419  \\
\hline
\end{tabular}
 \caption{\small Average multiplicities of
  secondary charged particles  at $P_{lab}=200$GeV/c }
\end{table}

The interactions of 200~GeV antiprotons on hydrogen,
argon and xenon were studied with the streamer chamber vertex
spectrometer \cite{De Marzo} at the CERN SPS.
Let us note, that the UrQMD model predicts low multiplicities
for all and negative charged particles in
$ \bar{p}p $-collisions.  At the same time, the model gives a larger
multiplicity of  particles for
$ \bar{p} $ interactions with heavy nuclei compared with experimental
data.
 As seen from Table~4, the UrQMD model calculations reproduce quite well
the experimental values of the average multiplicities of
particles emitted in a forward hemisphere.  The disagreement between the theory
and the experimental data on the average multiplicity of the backward
particles reaches 80$\%$ for  $\bar{p}Xe$-interactions.

\vspace{-0.5cm}
\begin{figure}[cbth]
\begin{center}
\psfig{file=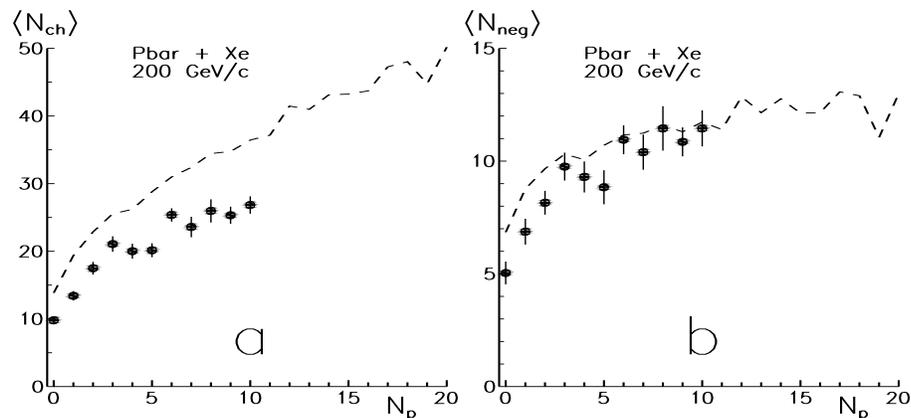,width=120mm,height=55mm,angle=0}
\vspace{-0.5cm}
\caption {{\small Correlations between protons and pions:
a) dependence of the average number of all charged particles,
$<N_{ch}>$, on the number of identified protons, $N_p$,
b) dependence
of the average number of negative charged particles, $<N_{neg}>$, on the
number of  protons.  Circles are  experimental data
\protect{\cite{De Marzo}}, dashed
lines are  UrQMD  calculations, respectively.}}
\label{fig10}
\end{center}
\end{figure}

\vspace{-0.5cm}
 Figs.~\ref{fig10}a,~~\ref{fig10}b show the average multiplicity of produced particles
$N_{ch}$ and negative particles  $ N_{neg}$ in  $ \bar{p}Xe $-events versus the number
$N_p$ of identified protons  with the momentum less than 600~MeV/c.
At high values of $N_p$ the average multiplicity
has a value by about three times more than that at $N_p=0$.
Fig.~\ref{fig10} illustrates the both average multiplicity
of charged particles and negative particles rise
with enhance of the identified proton number.
    The corresponding UrQMD model
 calculations reproduce the  experimental correlations qualitatively,
but there is a quantitative difference between
the simulations and the
data for dependence of the   average
multiplicity of all charged particles on the number of the identified
 protons. The number of  identified protons in simulations can be more than
 20. At the same time, the number $N_p$ in experiment reaches only  10.

\begin{figure}[cbth]
\begin{center}
\vspace{-0.8cm}
\psfig{file=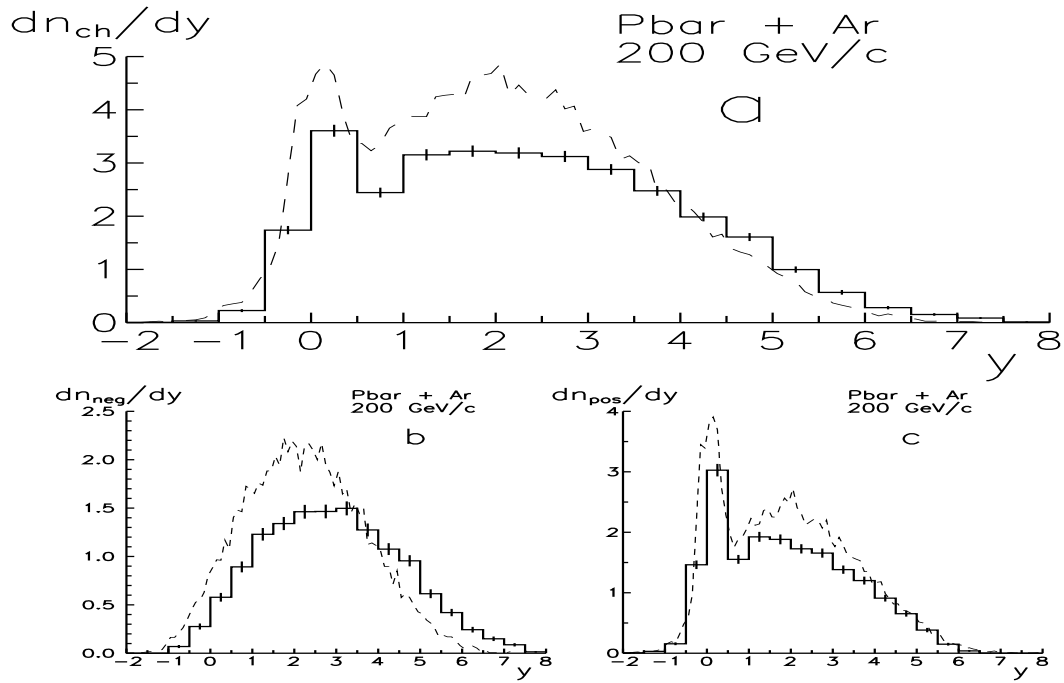,width=140mm,height=90mm,angle=0}
\vspace{-0.2cm}
\caption{{\small Rapidity distribution of charged particles
in $\bar{p}Ar$-interactions:
 a) all charged particles,
 b)negative charged particles,
 c) positive charged particles.
 Histograms are experimental data
 \protect{\cite{De Marzo}},  dashed
lines are UrQMD model calculations.}}
\label{fig11}
\end{center}
\end{figure}

\vspace{-0.7cm}
Rapidity distributions of  all,   negative an
 positive charged particles give more detail information about the
 accuracy of model predictions. In  Fig. \ref{fig11} we present the
 rapidity distributions  of  secondary  particles in
 $\bar{p}Ar$-interactions at 200~GeV/c.
In the rapidity distributions
of all and positive charged
particles, the  peak at $y \sim 0$ is associated with fast ejected target protons
or with dissociation of the protons.

  The following observations can be made:
\begin{itemize}
\item  The UrQMD model assumes dominant
production of all charged particles in the central rapidity region and
in the target fragmentation region.
\item The model gives  a shift of
the negative particle rapidity distribution to the target fragmentation
region.
\item  The model overestimates the multiplicity of positive
  charged particles in the central and the  target fragmentation
 regions.
 \end{itemize}

\begin{figure}[cbth]
\begin{center}
\vspace{-2cm}
\psfig{file=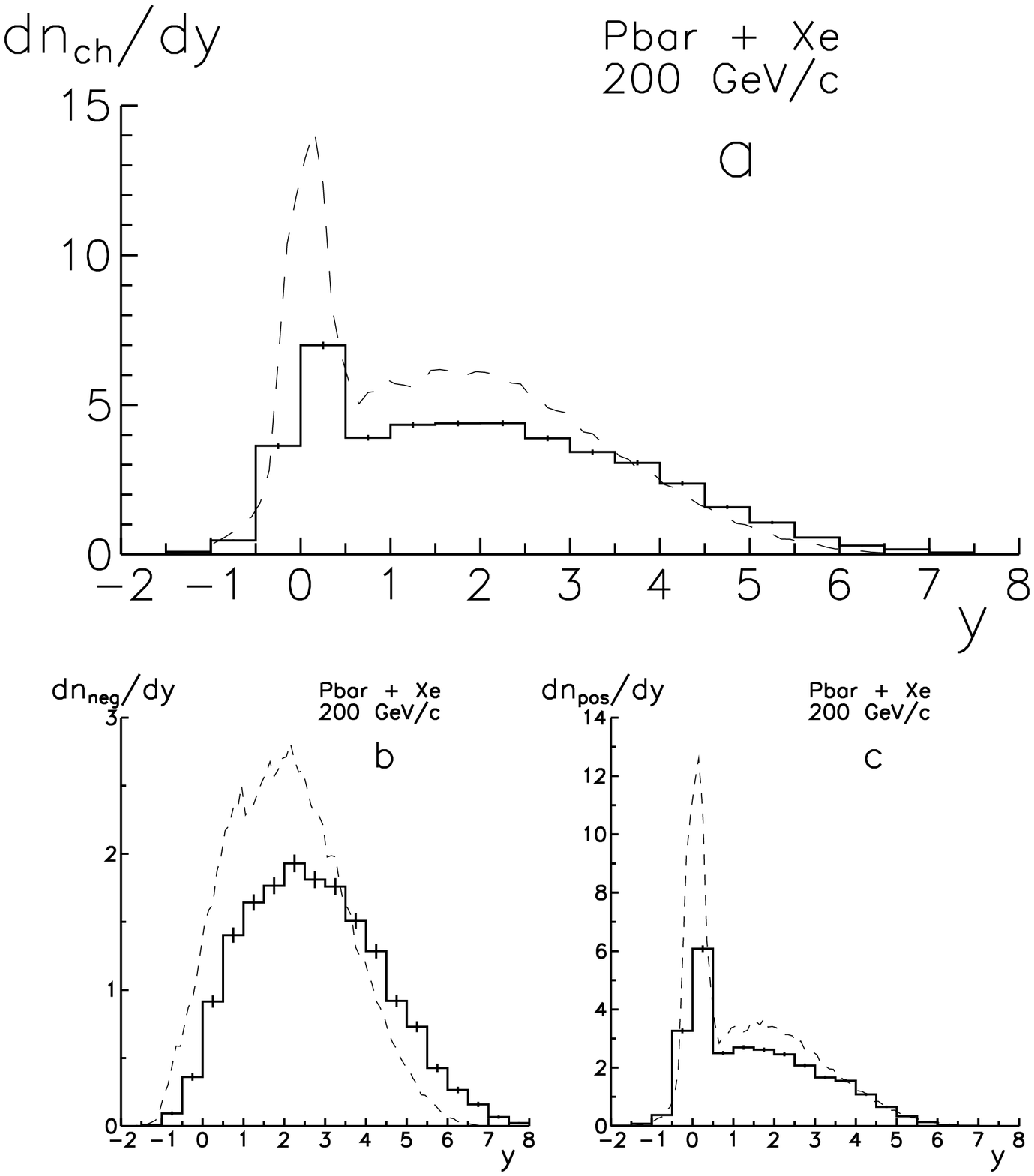,width=140mm,height=110mm,angle=0}
\vspace{-0.5cm}
\caption{{\small Rapidity distribution of charged particles in $\bar{p}Xe$-interactions:
 a) all charged particles, b) positive charged particles,
c) negative charged particles. Histograms are experimental
data \protect{\cite{De Marzo}}, points with dushed lines are
 UrQMD model calculations.}}
\label{fig12}
\psfig{file=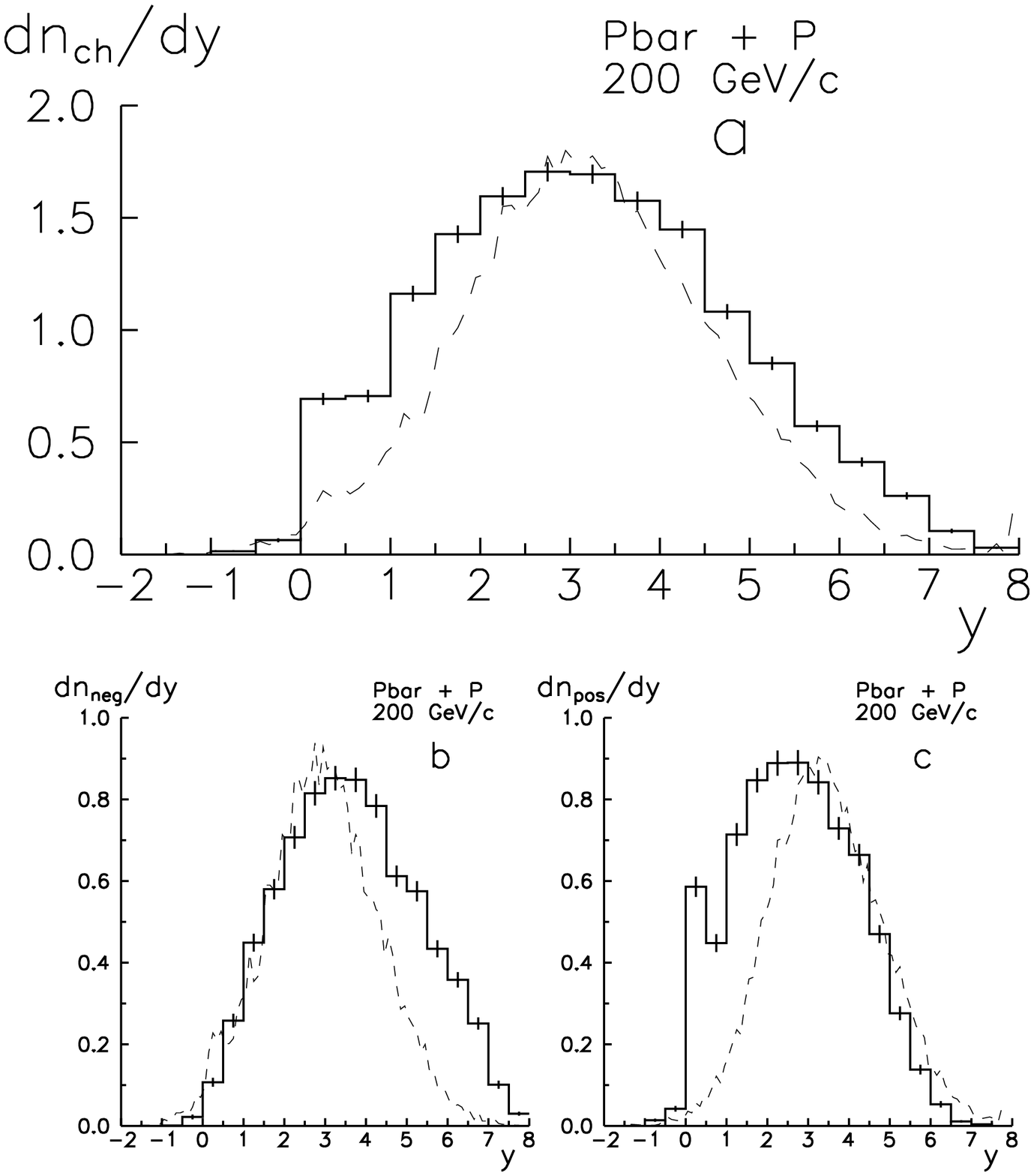,width=140mm,height=110mm,angle=0}
\vspace{-0.5cm}
\caption{{\small Rapidity distribution of charged particles in $\bar{p}p$-interactions:
 a) all charged particles, b) positive charged particles,
 c) negative charged particles. Histograms are experimental
 data \protect{\cite{De Marzo}},
 points with dushed lines are UrQMD model calculations.}}
\label{fig13}
\end{center}
\end{figure}

The same features  are also presented in  $\bar{p}Xe$-interactions
at momentum of the antiprotons at 200~GeV/c (Fig.~\ref{fig12}).
However, discrepancy between the calculated  and
experimental spectra in the central region and target fragmentation
regions is essentially larger than in $\bar{p}Ar$-interactions. It
 can be explained by effects of powerful cascading
 and a large number of produced resonances in the model $\bar{p}Ar$-simulations.

The data of $\bar{p}p$-interactions on rapidity distributions
(see Fig.~\ref{fig13}) show that the UrQMD approach does not describe
the target and projectile fragmentation regions. There is a clean
effect of  $\bar{p}$ leading and target proton leading in the
experiment.  However, the UrQMD approach assumes the central production of the
charged particles, and there is no  manifestation of the leading effect.
Thus, we can conclude that the UrQMD approach does not consider diffraction
dissociation quite correct.

\vspace{-0.3cm}
\begin{figure}[cbth]
\begin{center}
\vspace{-0.4cm}
\psfig{file=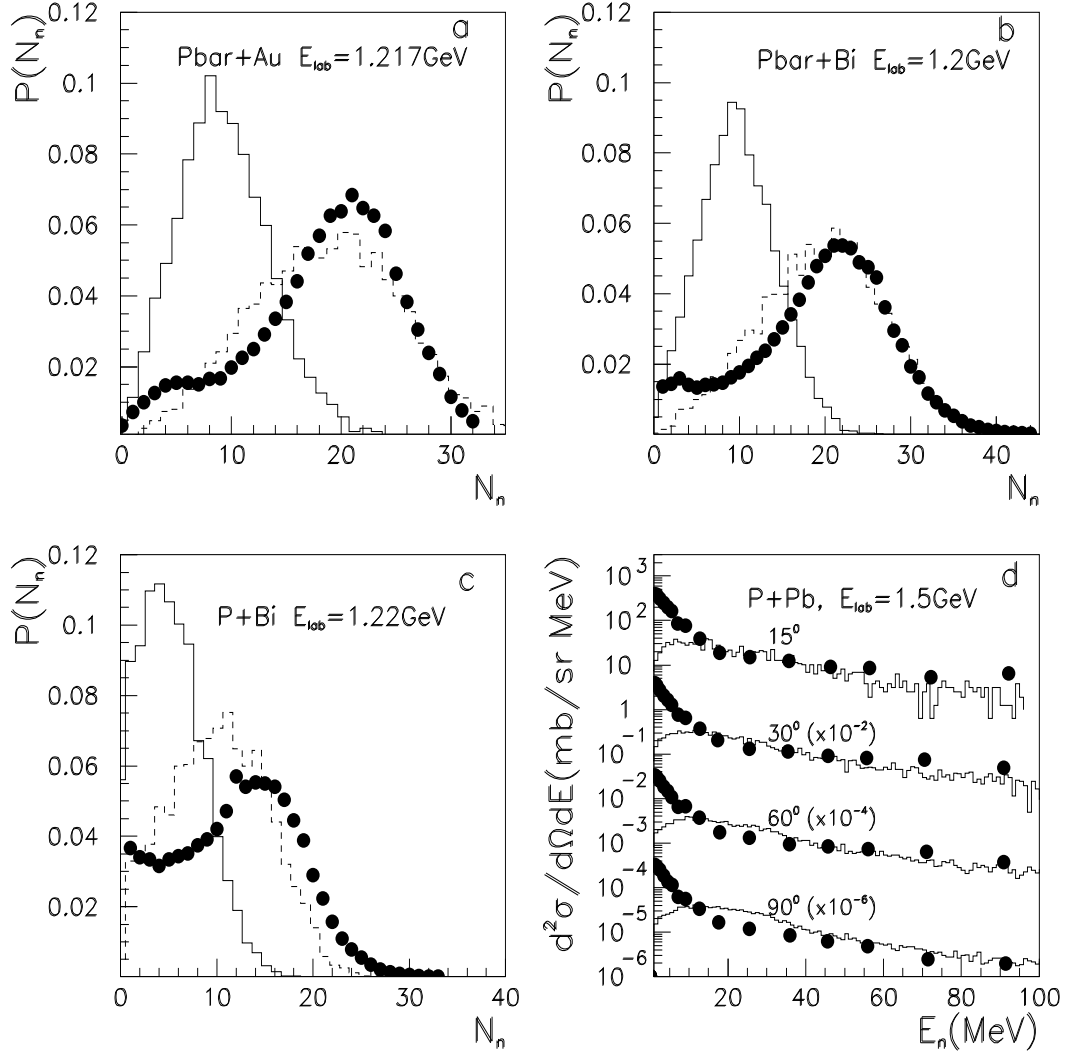,width=140mm,height=140mm,angle=0}
\vspace{-0.5cm}
\caption {{\small Multiplicity and energy distributions of produced  neutrons
in $\bar{p}/p~A$-interactions:
 a) multiplicity distribution of neutrons produced
 in $\bar{p}Au$-interactions,
 b) the same as in Fig. a for $\bar{p}Bi$-collisions,
 c) the same as in Fig. a for $pBi$-interactions.
 Points are experimental data \protect{\cite{Jahnke}},  solid  and dashed
lines are the UrQMD model calculations corrected and uncorrected
for neutron detection efficiency, respectively.
d) Energy distribution of neutrons produced at  angles
$15^{o}, 30^{o}, 60^{o}, 90^{o}$ in  $pPb$-interactions.
 Points are experimental data \protect{\cite{n_e}},  solid
lines are the UrQMD model calculations. }}
\label{neutr}
\end{center}
\end{figure}

\vspace{-0.5cm}
In order to complete the consideration of the UrQMD model predictions,
we have to analyse the products of nuclear residuals.
To estimate  radiation conditions of experiments, it is very important
to know the multiplicity and energy spectra of neutrons
produced by nucleus target  in $\bar{p}A$-interactions.
Thus, we start with the multiplicities of neutrons
produced in $\bar{p}Au, \bar{p}Bi, pBi $ -interactions at
the energy of projectile baryons 1.2~GeV.  Figs.
\ref{neutr}a, \ref{neutr}b, \ref{neutr}c give
 experimental data  \cite{Jahnke} and
calculation by the UrQMD model
of the neutron multiplicity distributions.
It is not clear from  Ref. \cite{Jahnke}, whether
the efficiency of the neutron detection by the Berlin Neutron Ball (BNB)
\cite{proc}
is taken  or not into account  in the experimental neutron multiplicity
distribution. Therefore, we have shown  both calculations uncorrected and
  corrected
for  neutron detection efficiency  (dashed and
solid lines in Figs. \ref{neutr}a, \ref{neutr}b, and \ref{neutr}c,
respectively).
The neutron detector BNB
counts the number of neutrons with efficiency $\sim 80\%$
in the evaporation energy regime. For higher neutron energy,
BNB detection efficiency drops to 30 -- 40 $\%$ for pre-equilibrium
neutron emission \cite{proc}.

As seen, the essential  missing of neutrons takes place
in the simulations corrected for the neutron efficiency.
The decrease of the produced neutron number is caused by
the absence  of the nucleus evaporation mechanism
in the UrQMD model.
To demonstrate this, we show  the energy spectra of neutrons
produced in $pPb$-collisions at the proton energy of 1.5~GeV.
In Fig.~\ref{neutr}d, the neutron energy distributions \cite{n_e}
at  the laboratory angles of emitted neutrons
$15^{o},30^{o} , 60^{o},
90^{o} $
are exposed. As seen from  Fig.~\ref{neutr}d, at the neutron
energy variation from 15~MeV to 1~MeV the calculated numbers of neutrons
(solid lines) decrease. At the same time, the experimental
spectra (points) grow up. Such slow neutrons are produced mainly
in the procceses of  target nucleus evaporation.
The underestimation of  the multiplicities of
the neutrons is at the level  of $\sim 15 - 25$.
We believe that it is  needed
to add the nucleus evaporation mechanism in the UrQMD model
for accurate estimation of the kinamatic characteristics  of  secondary
neutrons. It is also observed that the UrQMD model
predicts correctly the energy spectra of the pre-equilibrium,  fast
neutrons (see Fig.~\ref{neutr}d).

Summing up, we conclude that
the predictions of  the UrQMD model agree qualitatively
with the experimental data on $\bar{p}A$-interactions.
However, we believe that the accuracy of the UrQMD model predictions
is not sufficient to estimate   the background conditions for
the experiment.
The UrQMD model should be modified for a quantitatively
description of the experimental data on $\bar{p}A$-interactions.

~~~~The authors have benefited from instructive discussions with
V.V.~Uzhinsky, E.A.~Strokovsky  and J.~Ritman.

\end{document}